\documentclass[preprint]{aastex}

\usepackage{graphicx}
\usepackage{amsmath}
\usepackage{rotating}

\begin{document}

\title{Water Delivery and Giant Impacts in the `Grand Tack' Scenario}

\author{David P.~O'Brien}

\affil{Planetary Science Institute, 1700 E.~Ft.~Lowell, Suite 106, Tucson, AZ 85719, USA}

\email{obrien@psi.edu}

\author{Kevin J.~Walsh}

\affil{Department of Space Studies, Southwest Research Institute, 1050 Walnut Street, Suite 300\\Boulder, Colorado 80302, USA}

\author{Alessandro Morbidelli}

\affil{Universit\'e de Nice --- Sophia Antipolis, CNRS, Observatoire de la C\^ote d'Azur\\BP 4229, 06304 Nice Cedex 4, France}

\author{Sean N.~Raymond}

\affil{Universit\'e de Bordeaux, Observatoire Aquitain des Sciences de l'Univers, 2 Rue de l'Observatoire\\BP 89, F-33270 Floirac Cedex, France}
\affil{CNRS, UMR 5804, Laboratoire d'Astrophysique de Bordeaux, 2 Rue de l'Observatoire\\BP 89, F-33270 Floirac Cedex, France}

\author{Avi M.~Mandell}

\affil{NASA Goddard Space Flight Center, Code 693, Greenbelt, MD 20771, USA}

\affil{\vspace{0.2in} \it Published Open Access in Icarus, September 2014 (Volume 239, pp.~74-84)}
\affil{\it 5 Figures and 5 Tables at end of Paper}

\section*{ABSTRACT}

A new model for terrestrial planet formation \citep{Hansen2009ApJ, Walsh2011Nat} has explored accretion in a truncated protoplanetary disk, and found that such a configuration is able to reproduce the distribution of mass among the planets in the Solar System, especially the Earth/Mars mass ratio, which earlier simulations have generally not been able to match.  \citeauthor{Walsh2011Nat} tested a possible mechanism to truncate the disk---a two-stage, inward-then-outward migration of Jupiter and Saturn, as found in numerous hydrodynamical simulations of giant planet formation.  In addition to truncating the disk and producing a more realistic Earth/Mars mass ratio, the migration of the giant planets also populates the asteroid belt with two distinct populations of bodies---the inner belt is filled by bodies originating inside of 3 AU, and the outer belt is filled with bodies originating from between and beyond the giant planets (which are hereafter referred to as `primitive' bodies). 

One implication of the truncation mechanism proposed in  \citeauthor{Walsh2011Nat} is the scattering of primitive planetesimals onto planet-crossing orbits during the formation of the planets. We find here that the planets will accrete on order 1--2\% of their total mass from these bodies.  For an assumed value of 10\% for the water mass fraction of the primitive planetesimals, this model delivers a total amount of water comparable to that estimated to be on the Earth today. The radial distribution of the planetary masses and the dynamical excitation of their orbits are a good match to the observed system. However, we find that a truncated disk leads to formation timescales more rapid than suggested by radiometric chronometers. In particular, the last giant impact is typically earlier than 20 Myr, and a substantial amount of mass is accreted after that event. This is at odds with the dating of the Moon-forming impact and the estimated amount of mass accreted by Earth following that event.  However, 5 of the 27 planets larger than half an Earth mass formed in all simulations do experience large late impacts and subsequent accretion consistent with those constraints.

\noindent \textit{Keywords: Planetary formation; Planetary dynamics; Planets, migration}

\section{Introduction}

The timeline for the important processes of terrestrial planet formation extends from the condensation of the first solids $\sim$4.567--4.568 Gyr ago \citep{Amelin2002Sci, Bouvier2010NatGSci, Connelly2012Sci} until the end of heavy bombardment of the inner Solar System around 4.1--3.8 Gyr ago \citep{Tera1974EPSL, Chapman2007Icar, Bottke2012Nat}.  After the formation of the first solids, the gaseous solar nebula dissipated within $\sim$2--10 Myr \citep{Haisch2001ApJ, Kita2005ASPC}. During that short time, most planetesimals formed and the giant planets grew their cores and captured their substantial gas content.  The accretion of the terrestrial planets, however, did not finish until much later, around 30--100 Myr \citep{Raymond2006Icar, Raymond2009Icar, Kenyon2006bAJ, OBrien2006bIcar, Kleine2009GCA, Morbidelli2012AREPS}. 

In the `classical' picture of terrestrial planet formation, the terrestrial planet region is not affected by the giant planets other than through their distant gravitational perturbations.  The progression from planetesimals to terrestrial planets includes a few distinct phases of growth \citep[see][for a review]{Morbidelli2012AREPS}.  First, due to their dynamically cold orbits, planetesimals collide with low relative velocities, amenable to accretion, and grow into larger planetary embryos.  This stage of growth is termed ``Runaway Growth'' because the largest embryos accrete the fastest due to their larger cross section and enhanced gravitational focusing \citep[eg.][]{Greenberg1978Icar, Wetherill1989Icar, Kokubo1996Icar, Weidenschilling1997Icar}.  Eventually, the largest embryos begin to dynamically excite the orbits of the nearby planetesimals, increasing their relative velocities and making gravitational focusing less effective.  As the largest bodies slow their accretion, the smaller bodies are able to catch up.  This phase is termed ``Oligarchic Growth'' \citep{Kokubo1998Icar, Kokubo2000Icar}.  Once this stage nears completion and most large bodies attain relatively comparable masses, the smallest planetesimals will stay small because their relative velocities have increased and collisions between them will result in fragmentation rather than accretion.
 
The starting point for the final stage of terrestrial planet formation is a suite of planetary embryos in a sea of smaller remnant planetesimals. This has typically been modeled as an idealized bimodal mass distribution of planetesimals and embryos, with approximately equal mass in each population \citep{Chambers2001Icar, OBrien2006bIcar, Raymond2009Icar}.  This configuration was likely reached before the gaseous solar nebula dissipated, and the loss of the gas, which stabilized the embryos due to its damping effect, allowed the embryos to interact and excite each other onto crossing orbits.  Large collisions and mergers occurred over the next 30--100 Myr, building the terrestrial planets we have today.  In the classical scenario, the distribution of planetesimals and embryos would have extended into the asteroid belt region, and the current population of asteroids is all that remains following the dynamical depletion of that region by the scattering effects of planetary embryos and dynamical resonances with the giant planets \citep{Wetherill1992Icar, Petit2001Icar, OBrien2007Icar}.

A significant problem with terrestrial planet formation in the above scenario is that planets forming around the region of Mars ($\sim$1.5 AU) in the simulations are typically 5--10 times more massive than Mars \citep[see, eg. ][]{Wetherill1991LPSC,Chambers2001Icar,Raymond2004Icar,Raymond2006Icar,OBrien2006bIcar, Raymond2009Icar,Morishima2010Icar}.  The exploration of a wide range of parameter space by \citeauthor{Raymond2009Icar} and \citeauthor{Morishima2010Icar} highlighted the difficulty of the problem, finding that only cases with extreme and improbable orbits for the giant planets were able to consistently produce small Mars analogs.  Another case with extreme initial conditions was studied by \citet{Hansen2009ApJ}, in which all of the mass in the terrestrial planet region was initially concentrated between 0.7 and 1.0 AU.  The results were promising, with consistent formation of planetary systems with an Earth/Mars mass ratio matching the Solar System, and good matches for other dynamical metrics as well.  Essentially, the fact that the embryo disk was truncated at 1 AU allowed Mars to form as an embryo that was scattered from the edge of the disk and became isolated, accreting little additional mass.  While the initial conditions of \citet{Hansen2009ApJ} were \textit{ad-hoc} and not based on any physical model, the work was inspiring and led to a new exploration of possible terrestrial planet formation scenarios.

\subsection{The ``Grand Tack''}

A possible mechanism for truncating a disk of planetesimals and embryos in the terrestrial planet region was explored by \citet{Walsh2011Nat}, who modeled the effects of the migration of the giant planets while they were still embedded in the gas-rich protoplanetary disk.  Due to their disparate formation timescales, the giant planets would have been mostly formed before the final stages of terrestrial planet accretion began. Thus, any substantial radial migration of the giant planets could have a major impact on the final terrestrial planet system.  Giant planets of roughly a Jupiter mass or larger can clear a gap in the disk and migrate by a process referred to as ``type-II'' migration, in which the planet moves inwards, following the viscous evolution of the disk \citep{Lin1986ApJ, Kley2012ARAA}.  However, the presence of a second giant planet of comparable but smaller mass exterior to the first one significantly changes the gap profile in the disk, and can halt and reverse the planet's inward migration \citep{Masset2001MNRAS, Morbidelli2007bIcar, Pierens2008AA, Pierens2011AA, DAngelo2012ApJ}. This migration reversal allows for a possible Solar System evolution in which Jupiter migrated inwards to the terrestrial planet region before being `caught' by Saturn, at which point their respective gaps merge into a common gap in the protoplanetary disk, causing them to migrate outward to roughly their current locations.

If Jupiter could have migrated into the inner Solar System and then back out, it would have had dramatic effects on the formation of the terrestrial planets. \citet{Walsh2011Nat} proposed that an inward-then-outward migration scenario may have truncated the surface density profile of the planetesimal disk to give roughly the initial conditions of \citet{Hansen2009ApJ}, which would help to form a small Mars.  The simulations of \citeauthor{Walsh2011Nat} explicitly accounted for the migration of the giant planets and their effect on the planetesimal disk, and used a somewhat different initial mass distribution than \citeauthor{Hansen2009ApJ} (with a bimodal mass distribution of planetesimals and embryos, rather than just a single population of embryos).  Despite these differences, both works found distributions of mass versus semimajor axis of the simulated planetary systems, as well as dynamical excitation, that were a good match to the Solar System.  In particular, inward migration of Jupiter to $\sim$1.5 AU would give an edge to the disk at $\sim$1 AU, which \citeauthor{Hansen2009ApJ} showed was ideal for forming Mars as a scattered and isolated planetary embryo.

One major constraint on such a scenario, however, is the existence and properties of the asteroid belt, which would have been profoundly affected by Jupiter migrating through it.  In modeling this scenario, dubbed the ``Grand Tack'', \citet{Walsh2011Nat} found that the asteroid belt, rather than being destroyed, would actually be scattered into place from two separate source populations. First, some planetesimals initially interior to Jupiter are scattered outward by Jupiter's inward migration and then back inward by Jupiter's outward migration. These planetesimals that make the round trip are preferentially scattered back to the inner region of today's asteroid belt. Second, some planetesimals that start between or beyond the orbits of the giant planets can be scattered into the asteroid belt, and those that follow this evolution typically land in the outer belt.

\citet{Walsh2011Nat} then assume that the planetesimals from inside the orbit of Jupiter are similar to the broad class of ``S-type''  asteroids, and those from outside the orbit of Jupiter are primitive asteroids, similar to the diverse class of ``C-type'' asteroids. The first reason for this assumption is the distribution of those bodies in the asteroid belt today and the dichotomy in their physical properties, where water-poor asteroids predominate in the inner region and water-rich bodies are more abundant in the outer region \citep{Gradie1982Sci, MotheDiniz2003Icar}. Second, it is typically believed that Jupiter formed near the water-condensation front (the `snow line') of the solar nebula and thus bodies forming outside the orbit of Jupiter would preferentially be water-rich. Thus, with these two broad classes of asteroids being drawn from distinct and separate reservoirs, a major success of \citet{Walsh2011Nat} was to reproduce, to first-order, the asteroid belt with the appropriate total mass, orbital and taxonomic distributions.

\citet{Walsh2011Nat} modeled the scattering of primitive C-type planetesimals from beyond Jupiter, some of which end on orbits in the asteroid belt. The same scattering mechanism also places many primitive planetesimals on higher-eccentricity orbits with perihelia below $\sim$1 AU. While \citeauthor{Walsh2011Nat} calculated the total mass of water-carrying material that was crossing the terrestrial planet region, they did not explicitly model the accretion of primitive planetesimals onto the terrestrial planets. Here we run planet formation simulations where the scattered population of primitive planetesimals is included, with the goal of quantifying the water-delivery process implied in this scenario. The primary focus is to assess the total amount of water delivered, the timing of the delivery, and the nature of the collisions responsible for the accretion of water-bearing material. There are also additional aspects of the terrestrial planet formation process that were not analyzed in detail in \citet{Walsh2011Nat} that we will include here, such as chronological constraints related to the accretion timescales of the terrestrial planets (in particular the last giant impact event that could be responsible for forming the Moon), and the planetesimal accretion following the last giant impact, which is relevant to the Earth's ``late veneer'' of highly siderophile elements.

\section{Simulations}
\label{simulations}

The simulations here are an extension of those presented in \citet{Walsh2011Nat}, who first proposed and modeled the Grand Tack scenario.  In their simulations, Jupiter and Saturn first migrate inwards for 100 kyr then outwards for 500 kyr, with Jupiter reversing its migration at 1.5 AU. Their end locations coincide with the initial conditions in the \textit{Nice Model} \citep[eg.][]{Morbidelli2007AJ,Levison2011}, such that Jupiter (a=5.285 AU) and Saturn (a=7.106 AU) are in a mutual mean-motion resonance, in this case the 3:2 resonance.

\citeauthor{Walsh2011Nat} primarily modeled how the migration of the giant planets could set up the initial conditions necessary for terrestrial planet formation from a narrow annulus, and what effect that migration would have on the asteroid belt.  They also performed a set of terrestrial planet formation simulations starting from the mass distribution at the end of the Grand Tack, at time 600 kyr. To limit the computational time required, only the material that was initially inside the orbit of Jupiter and that was inside of 2~AU  at the end of the Grand Tack was included, as it dominated the total mass.  Hence, none of the potential water-bearing planetesimals from beyond the snow line were included in the terrestrial planet formation simulations.  Here, in order to address water delivery issues we also include the planetesimals that started outside the orbit of Jupiter and were subsequently scattered inward. 

\citeauthor{Walsh2011Nat} used two different initial embryo and planetesimal distributions interior to 3 AU.  At model time t=0, which is after planetesimals and embryos have formed but prior to the Grand Tack migration, their SA151--SA154 simulations started with 37 1/2-Mars-mass embryos from 0.7 to 3 AU and the SA161--SA164 simulations started with 74 1/4-Mars-mass embryos over the same range.  These simulations also included 727 planetesimals spanning the same semimajor axis range, which were termed the `S-Type' planetesimals, each with a mass of 0.00255 $\mathrm{M_{\earth}}$ (about 1/40 of a Mars mass).  There was thus an equivalent amount of mass in the embryo and planetesimal populations in the simulations.  The total mass present in solids is consistent with a minimum-mass solar nebula (MMSN) distribution \citep[eg.][]{Hayashi1981PTPS}.  Since nebular gas is present during the first 600 kyr of the simulations in order to drive the giant planet migration, these simulations also included the effect of gas drag on the planetesimals and the effects of tidal eccentricity damping on the embryos.  The gas drag calculations assume a planetesimal diameter of 100 km, consistent with recent models of planetesimal formation \citep{Johansen2007Nat, Cuzzi2008ApJ}.  While we do not model it explicitly, a plausible origin of the assumed inner edge of the embryo and planetesimal distribution at 0.7 AU is Type-1 migration due to the tidal interaction of embryos with the primordial gas disk. \citet{Ida2008ApJ} found that Type-1 migration can efficiently clear solid material from the inner regions of a disk. 

After 600 kyr, the migration of the giant planets has shepherded material inwards (mainly by the combined effects of resonance trapping and gas drag) and compressed much of the mass initially present inside of 3 AU into a narrow annulus between $\sim$0.7 and 1 AU, containing roughly 2 $\mathrm{M_{\earth}}$ of material.  This inward shepherding of material roughly doubles the amount of mass present in the 0.7--1 AU range compared to the amount at $t=0$, achieving a high mass concentration in this region from MMSN-like starting conditions.  On average, 10 embryos are present after 600 kyr in the SA151--SA154 simulations, and 22 are present in the SA161--SA164 simulations (a significant burst of accretion occurs during this time, such that many of the embryos have grown in mass).  An average of 390 of the S-Type planetesimals also remain after 600 kyr. These conditions after 600 kyr are the starting point for the terrestrial planet formation simulations that we perform here.  We note that this does not correspond to 600 kyr after CAI formation; Our model time t=0 is after planetesimals and embryos have formed, and that process likely occurred over several Myr \citep[eg.][]{Scott2006Icar}.

The primitive ('C-Type') planetesimals in the \citet{Walsh2011Nat} simulations were modeled to come from two different populations:~Those from the regions in-between the giant planets' orbits, hereafter denoted the ``belts'', and 
those beyond the giant planets hereafter denoted the ``disk''.  Many of these planetesimals are scattered into the asteroid belt and onto terrestrial planet-crossing orbits by the migration of the giant planets during the first 600 kyr.  They were treated as massless particles since their scattering and implantation efficiencies were not yet understood.  To determine the amount of water delivered to the planets, these primitive planetesimals must be assigned a mass, and we do this based on the following considerations.

\citet{Walsh2011Nat} estimate that $1.3 \times 10^{-3} \ \mathrm{M_{\earth}}$ of inner-Solar System material (`S-Type') is implanted into the asteroid belt in their simulations. This total mass of asteroids is a direct result of the scattering efficiency from the inner disk of material, whose total mass is set by the requirement that $\sim$2 Earth masses must remain to build the terrestrial planets.  The main belt has approximately three times more C-Type bodies than S-Type \citep{MotheDiniz2003Icar}, which means that approximately $3.9 \times 10^{-3} \ \mathrm{M_{\earth}}$ of C-Type material must be implanted into the main belt.

\citeauthor{Walsh2011Nat} find that on average for every primitive planetesimal that is implanted into the asteroid belt from the ``belts'' population, 17.6 planetesimals will attain q$<$1.5 AU and 20.3 planetesimals will attain q$<$2.0 AU.  For the ``disk'' population, the corresponding factors are 28.0 with q$<$1.5 AU and 44.5 with q$<$2.0 AU.  We can find the total masses of the q$<$1.5 AU and q$<$2.0 AU belts and disk populations using these factors and the total implanted mass of C-type material ($3.9 \times 10^{-3} \ \mathrm{M_{\earth}}$) calculated above.  The total mass of material from each population with q$<$1.5 AU is therefore $17.6 \times (3.9 \times 10^{-3} \ \mathrm{M_{\earth}})$ = 0.0687 $\mathrm{M_{\earth}}$ (belts) or $28 \times (3.9 \times 10^{-3} \ \mathrm{M_{\earth}})$ = 0.109 $\mathrm{M_{\earth}}$ (disk).  The corresponding values for q$<$2.0 AU are 0.0793 $\mathrm{M_{\earth}}$ and 0.174 $\mathrm{M_{\earth}}$.

We take 285 of the ``belt'' particles from \citeauthor{Walsh2011Nat} as representative of the orbital distribution having $q<1.5$ AU, and knowing the total mass for the population (calculated above) we can assign a mass to each particle. In the case where the belts are the only source of C-type planetesimals, this gives an individual particle mass of 0.0687 $\mathrm{M_{\earth}}/285 = 2.4105 \times 10^{-4} \ \mathrm{M_{\earth}}$.  For q$<$2 AU, a subset of 325 of the belt particles is used, with the same individual mass. Similarly, in the case where the ``disk'' is the only source of C-type planetesimals, we consider 278 disk particles with $q<$1.5 AU and 442 with $q<$2.0~AU,  yielding a mass of 0.109 $\mathrm{M_{\earth}}/278 = 3.9209 \times 10^{-4} \ \mathrm{M_{\earth}}$ each.  The orbital elements $a$, $e$ and $i$ for these particles are taken directly from the simulations of \citeauthor{Walsh2011Nat}, and the angular elements $\Omega$, $\omega$ and $M$ are randomly assigned.

The calculation of the masses of the primitive belt and disk planetesimals above assumes that all of the mass of the C-Type asteroids in the asteroid belt comes either entirely from the belt population or entirely from the disk population.  Thus, although we include both disk and belt planetesimals in each simulation, we will calculate the final amount of primitive material accreted by the terrestrial planets by considering disk planetesimals and belt planetesimals separately, and not the sum of the two.  In reality, because the primitive planetesimal population should be a combination of those coming from the belts and from the disk, the actual amount of primitive material accreted by the planet will be bracketed by the two values that we compute.  

\begin{figure}[]
\centering
\includegraphics[width=6.5in]{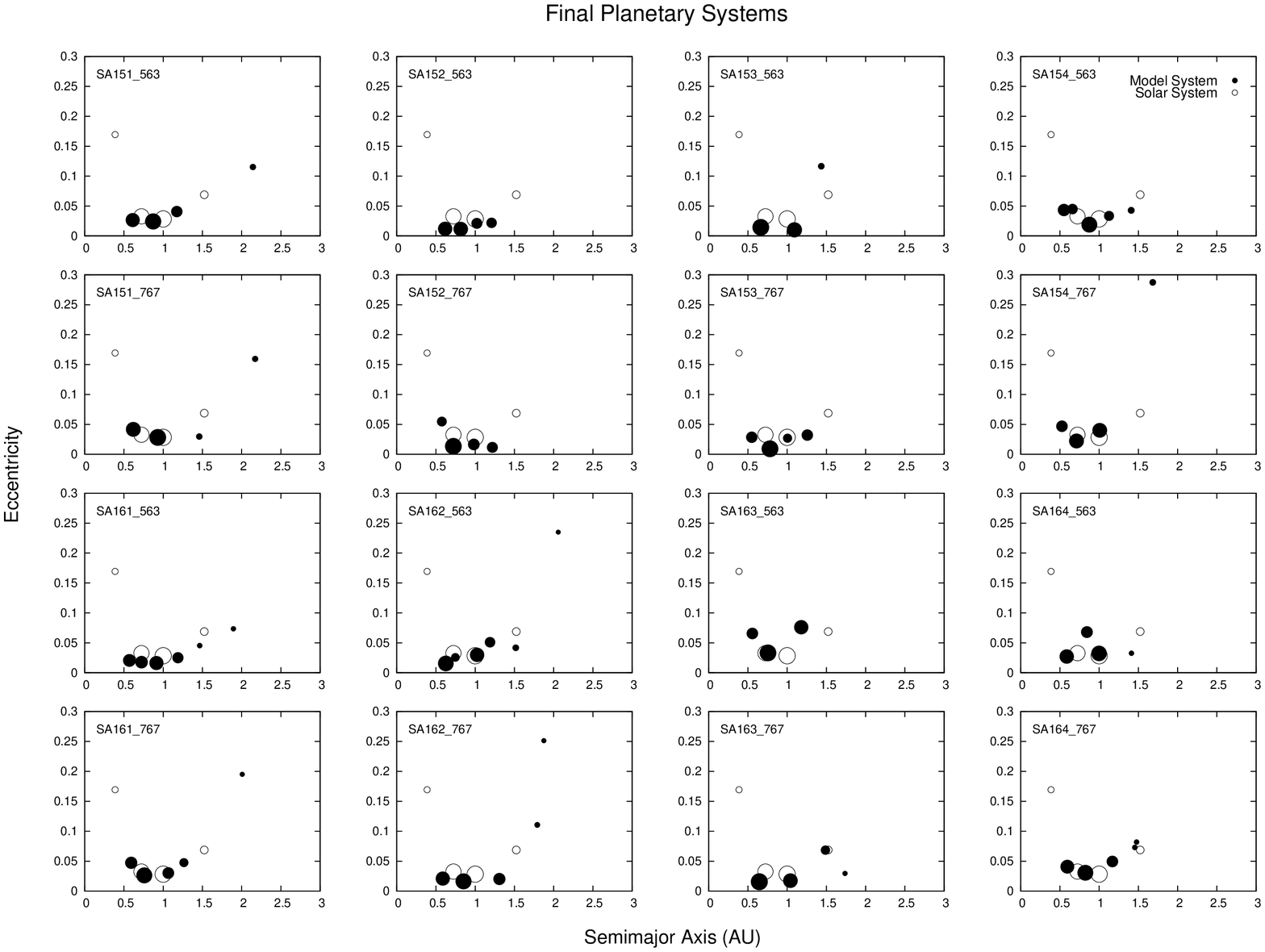}
\figcaption{Final planetary systems for all simulations.  The SA15* simulations start with 1/2-Mars-mass embryos and the SA16* simulations start with 1/4-Mars-mass embryos.  The *563 simulations include all primitive planetesimals with perihelion $q<1.5$ AU and the *767 simulations include all primitive planetesimals with $q<2$ AU.  Orbital elements are averaged over the last 1 Myr of the simulations. Open circles show the current orbits of the terrestrial planets in the Solar System. \label{systemsfig}}
\end{figure}

\begin{figure}[]
\centering
\includegraphics[width=5.5in]{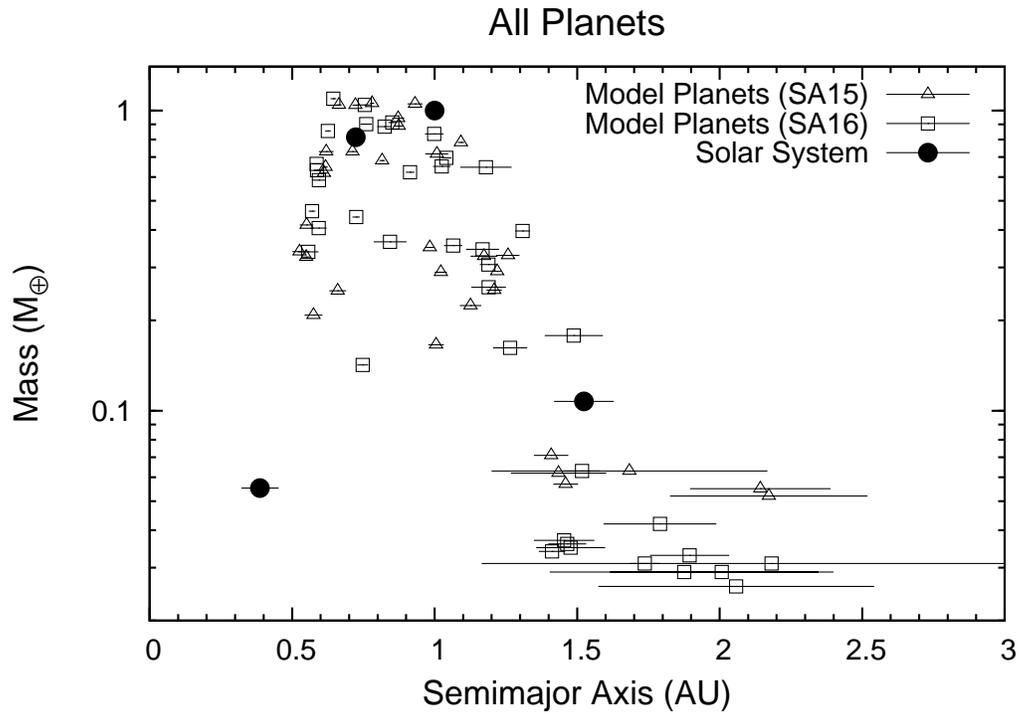}
\figcaption{Mass distribution of planets from all simulations.  The SA15 simulations, shown as triangles, start with 1/2-Mars-mass embryos and the SA16 simulations, shown as squares, start with 1/4-Mars-mass embryos.  Horizontal lines show the range of perihelion-aphelion.  Solid circles show the actual terrestrial planets. \label{massdistfig}}
\end{figure}

For the initial conditions of the simulations we run here, we take the results of the \citet{Walsh2011Nat} simulations SA151--154 and SA161--164 after 600 kyr for the embryos and S-Type planetesimals, and for each of them we use a case that includes primitive C-Type planetesimals with q$<$1.5 AU and another one with q$<$2 AU, giving 16 simulations in total.  Simulations with q$<$1.5 AU are noted as *563 (eg.~SA151\_563), where 563 is the total number of belt and disk primitive planetesimals included, and those with a cutoff of q$<$2.0 AU are likewise denoted as *767.  The combined distributions are integrated for 150 Myr using the SyMBA numerical integrator \citep{Duncan1998AJ} with a timestep of 7 days (sufficient to accurately integrate orbits beyond $\sim$0.5 AU).  The final terrestrial planet systems are shown in Fig.~\ref{systemsfig}, and the radial distribution of planet masses is plotted in Fig.~\ref{massdistfig}.  These will be discussed in further detail in the subsequent sections.

\section{Analysis}

Here we look in detail at the final planetary systems and the evolution of individual planets in our simulations (those shown in Fig.~\ref{systemsfig}).  First, the basic quantitative metrics of the mass distribution and orbits can be compared with the Solar System and those from previous simulations.  We then look in more detail at the timescales of accretion compared to geochemical constraints, the accretion of primitive planetesimals and the implications for water delivery to Earth, and the implications for Moon-forming impacts.

\subsection{Dynamical Metrics for the Planetary Systems}

A quantitative metric for the distribution of planet mass as a function of semimajor axis is the radial mass concentration (RMC) statistic \citep{Chambers2001Icar} that is defined as
 
\begin{equation}
S_c = \mathrm{max} \left(\frac{\sum_j m_j}{\sum_j m_j [\mathrm{log_{10}}(a/a_j)]^2}\right)
\end{equation}

\noindent where $m_j$ and $a_j$ are the mass and semi-major axis of planet $j$, and $S_c$ is the maximum value of the function in brackets, evaluated over all $a$.  It has a value of 89.9 for the Solar System, and larger values indicate a higher degree of mass concentration in a narrow region (eg.~Earth and Venus contain most of the mass of the terrestrial planets and are fairly close to one-another, giving a large value of $S_c$).

We calculate this value (and the angular momentum deficit in the following paragraphs) using only planets and embryos with an orbit interior to 2~AU.  We expect that anything beyond 2~AU will be lost during later giant planet migration and ejected from the Solar System.  This is not precisely the same qualification as used by \citet{Walsh2011Nat}, where they considered as a planet all bodies with a final mass larger than 0.03 $M_{\earth}$.
The values we give here for \citeauthor{Walsh2011Nat} are re-calculated from their simulations using the current criterion for direct comparison. 

The $S_c$ values for this suite of simulations are reported in Table \ref{amdtable}, with median values of 88.6 for the SA15 simulations and 71.2 for the SA16 simulations. These values are comparable to that for the Solar System (89.9), and as shown in Fig.~\ref{massdistfig}, the radial mass distribution of the final planets in our systems follows the same general trend as in the Solar System, with most of the mass concentrated in planets within or near $\sim$0.7--1 AU.  Our simulations do succeed in producing planets with roughly the location and mass of Mars.  However, there are no good Mercury analogues.

As expected, the $S_c$ values here are comparable to the median values for the \citet{Walsh2011Nat} SA15 and SA16 simulations (88.9 and 71.1, respectively), which used the same initial conditions as our simulations but without the primitive planetesimals.  Similarly, \citet{Hansen2009ApJ} found a median of 86.1 for all 38 simulations he ran, with minimal differences between those with Jupiter starting at 0 or 5 Myr.  For comparison, simulations without a truncated disk, with an initial mass distribution extending from $\sim$0.5 to 2--4 AU \citep[eg.][]{Chambers1998Icar, Chambers2001Icar, OBrien2006bIcar, Raymond2009Icar}, generally find $S_c$ values less than $\sim$50.  This implies much less radial mass concentration of the planets, and a worse match to the Solar System, than the simulations here and in \citet{Walsh2011Nat} and \citet{Hansen2009ApJ}.

The excitation of a planetary system can be quantified as the relative or normalized angular momentum deficit (hereafter abbreviated just as AMD), and is denoted $S_d$:

\begin{equation}
S_d = \frac{\sum_j m_j \sqrt{a_j(1-e_j^2)} \cos i_j - \sum_j m_j \sqrt{a_j}}{\sum_j m_j \sqrt{a_j}}
\end{equation}

\noindent \citep{Laskar1997AA, Chambers2001Icar}, where $m_j$, $a_j$, $e_j$, and $i_j$ are the mass, semi-major axis, eccentricity, and inclination of planet $j$.  This quantity is the fractional difference between the Z-component of the angular momentum of the system and the angular momentum of the system if all $e$ and $i$ were zero.  More negative values of $S_d$ imply a higher dynamical excitation.  The angular momentum deficits of all systems formed in our simulations are shown in Table \ref{amdtable}, calculated using orbital elements averaged over the last 1 Myr of the simulations.  The angular momentum deficit of the actual terrestrial planets is -0.0018 when averaged over million-year timescales \citep{Chambers2001Icar}.

It should be noted that the terrestrial planets may have had their orbits dynamically excited during the giant planet instability associated with the late heavy bombardment (LHB) \citep{Agnor2012ApJ,Brasser2013MNRAS}. \citeauthor{Brasser2013MNRAS} suggest that a post-formation AMD value on order of -0.001 would have a good probability of being excited to today's value during the LHB.  \citet{Brasser2013MNRAS} found that Mars' orbit, however, would likely suffer minimal excitation due to these dynamical processes, and was very likely on an orbit similar to its current orbit immediately following its formation.  Fig.~\ref{systemsfig} suggests that this is often the case in our simulations.  Mars-like planets are often scattered out of the truncated disk and isolated from substantial subsequent accretion, with a more excited orbit than for the planets that form from the material still left in the truncated disk.

The two sets of simulations presented here have median AMD values of -0.00061 and -0.0017 for SA15 and SA16 respectively. These simulations compare reasonably well with the AMD of the terrestrial planets of the Solar System, with the both sets of simulations spanning a range of AMD values that includes the \citet{Brasser2013MNRAS} primordial AMD estimate as well as the current Solar System value. In comparison, the median AMD values re-calculated for the simulations in \citet{Walsh2011Nat} for the terrestrial planet formation simulations without the primitive planetesimals are -0.00089 and -0.0022 for the SA15 and SA16 sets of simulations, respectively. The entire suite of simulations presented in \citet{Hansen2009ApJ} had a median value of -0.00195, with minimal differences for the different starting times of Jupiter. 

These results are consistent with the finding that dynamical friction from small planetesimals yields a low AMD value compared to simulations that lack small planetesimals, as discussed in \citet{OBrien2006bIcar} and \citet{Morishima2008ApJ}.  For comparison, the \citet{Chambers1998Icar} Model C simulations, each consisting of at most 50 large embryos extending out to 4 AU, have a median AMD of -0.033, and the \citet{Chambers2001Icar} Simulations 21--24, each consisting of a bimodal distribution of 150 bodies, have a median AMD of -0.0050.  The \citet{OBrien2006bIcar} simulations, which included a population of $\sim$1000 planetesimals, gave median AMD values of -0.0030 and -0.0010 for the two sets of simulations they ran, comparable to the values obtained here and in \citet{Walsh2011Nat}.

Dynamical friction may also explain why the median AMD for our SA15 simulations is about a factor of 3 lower than for SA16 simulations (although the range in values is quite large and overlapping for both; see Table 1). The SA15 simulations begin with embryos that are twice as large and twice as widely spaced as in the SA16 simulations. Thus, owing to larger mutual separations (in terms of Hill Radii), the SA15 embryos experienced less mutual gravitational stirring. At the same time, they also experience more effective dynamical friction due to the larger mass ratio between the embryos and planetesimals. These two factors contribute to the different measured values of AMD for the two suites of simulations.

Interestingly, \citet{Hansen2009ApJ} started with 400 embryos, each with $5\times10^{-3} \ M_{\earth}$, and no planetesimals, and one might expect high AMD values to result from the lack of dynamical friction. Instead they measure AMD values similar to those in our simulations, which had hundreds of planetesimals.  A likely explanation is that given the close packing of the initial embryos in their simulations, where the Hill Radii are initially overlapping for most bodies, there would be a large number of mergers early on and some larger embryos would quickly form, leading to a bi-modal distribution of bodies not long after t=0.  This would result in dynamical friction on the large bodies that form, since the mass of the initial particles in \citet{Hansen2009ApJ} is only 2 times larger than the planetesimals in the \citep{Walsh2011Nat} simulations and those presented here.

\subsection{Accretion Timescales, Last Giant Impacts, and Late Veneer}

Radiometric dating provides valuable constraints on the accretion timescales of the terrestrial planets.  The emerging data suggests substantially different timescales of accretion for different planets, with estimates of 30--100 Myr for Earth vs.~2--10 Myr for Mars \citep[eg.][]{Halliday2000EPSL, Halliday2006MESS2, Kleine2009GCA, Nimmo2007Icar, Dauphas2011Nat}. The very short timescale for the accretion of Mars suggests that it may have formed as a planetary embryo from runaway and oligarchic growth, but essentially did not participate in a later giant impact phase of planetary growth \citep[]{Dauphas2011Nat}.

In Table \ref{growthtimes}, the accretion timescales for all planets consisting of two or more embryos in our simulations is broken down into results for different final mass ranges.  The times to grow to 50\% and 90\% of the final mass are given.  For nearly all planets, accretion to 50\% mass is complete within a few Myr.  The timescale to accrete to 90\% of final mass is in the range of $\sim$20--30 Myr, with the largest planets ($M>0.75 \ \mathrm{M_{\earth}}$) accreting somewhat more quickly than the smaller planets. 

For Mars-mass planets, we find that the 90\% accretion time is much too long to match the radiometric chronometers mentioned above. One possible explanation is that the late large impact(s) that bring the planet to a Mars mass do not reset the Hf-W chronometer that is generally used as a measure of accretion timescale \citep{Morishima2013EPSL}.  Another possibility that is perhaps more likely in the Grand Tack scenario is that the embryos used in our simulations are too small.  The work of \citet{Hansen2009ApJ} and \citet{Walsh2011Nat} (also the simulations here) suggest that Mars was scattered out of the annulus of accreting bodies, leaving it `stranded' on a somewhat dynamically excited orbit.  Some simulations result in a single stranded embryo near 1.5~AU, which would satisfy the chronological constraints for Mars because the accretion timescales of those bodies would have been set by the time of the disk truncation.  If these embryos were somewhat larger (closer to a Mars mass), that could provide a possible solution.  This is being explored further in separate and ongoing work \citep{Jacobson2014Sub}.

For Earth-mass planets, the accretion timescales found here are faster than those typically reported for geochemical timescales \citep{Kleine2009GCA}. The timescale found in this work for the large planets is similar to that found in \citet{Hansen2009ApJ} ($t_{90}$=17 Myr), but it is much shorter than those found in the \citet{OBrien2006bIcar} simulations ($t_{90}$=65--70 Myr and 35--45 Myr, depending on the orbital excitation of Jupiter). The short timescales in the truncated disk simulations are due to the confinement of the embryos in the annulus, which enhances the mutual collision probabilities. In extended disk simulations, embryos initially at larger semi-major axis are more likely to accrete at later times because of longer orbital periods \citep{Raymond2006Icar}. Moreover, in \citet{Hansen2009ApJ}, rapid mutual accretion of embryos begins immediately due to overlapping Hill Radii, and in \citet{Walsh2011Nat} embryo-embryo collisions are triggered by the strong perturbations from Jupiter's migration. In the extended disk simulations, however, it takes much longer for the embryos to become unstable and start colliding with each other \citep{Chambers1998Icar,Morishima2010Icar}.

Another way to quantify accretion timescales is to use the time of the last major impact by a body of one embryo mass or larger. This is typically closely related to the accretion timescales, but for the case of the Earth is better constrained than $t_{90}$ because it is related to the formation of the Moon.  Another related constraint is the amount of accretion onto the planets following the last giant impact, often termed the ``late veneer'' in the case of the Earth.  Since that material would be accreted after major differentiation has ended, the highly-siderophile elements (HSEs) delivered in the late veneer would mostly remain in the mantle.  The HSE abundances in the crust and mantle of the Earth can constrain the total amount of material that was delivered during the late veneer, and has been estimated to be less than $\sim$1\% of the total mass of the Earth \citep{Drake2002Nat}.  \citet{Jacobson2014Nat} show that there is an inverse correlation between the amount of mass accreted after the last giant impact and the time of the last giant impact.

In Table \ref{lastimpactstbl}, we track the time of the last impact, $t_{imp}$, the mass $M_{imp}$ and approach velocity $V_{\infty}$ of that impactor, and the total accreted mass $M_{veneer}$ that arrives after this impact (as a percent of final planet mass).  The mass and velocity of the final impactor on the Earth are constrained by simulations of the formation of the Moon.  \citet{Canup2004Icar} found that the required impact conditions for reproducing the Earth-Moon system involved an impactor with a mass of 0.11--0.14 $\mathrm{M_{\earth}}$ impacting at a shallow angle (rather than head-on) with $V_{\infty} < $ 4 km/s. Newer simulations suggest that range of acceptable impact conditions may in fact be much larger, as \citet{Cuk2012Sci}, \citet{Canup2012}, and \citet{Reufer2012} found a range of impactor masses five times as broad with possible impact velocities up to 1.3 $\mathrm{V_{esc}}$.  Our median $V_{\infty}$ values from Table \ref{lastimpactstbl} are generally comparable to or less than the 4 km/s value found by \citeauthor{Canup2004Icar}.  For planets close to an Earth mass, the median impactor masses are somewhat less than the 0.11--0.14 $M_{\earth}$ value found by \citeauthor{Canup2004Icar}, although a significant number of impactors do fall into that size range.

A more difficult constraint is the timing of the of the final giant impact. 
For the Earth-Moon system, this can be constrained by Hf-W dating to be at least $\sim$50 Myr \citep[eg.][]{Touboul2007Nat, Kleine2009GCA} (However, see \citet{Konig2011} for a discussion of possible earlier Moon formation times).  The median $t_{imp}$ in Table \ref{lastimpactstbl} is substantially lower than this, with $t_{imp}$ larger than 50 Myr being very rare.  In comparison, the median value of $t_{imp}$ in the simulations of \citet{OBrien2006bIcar} with a dynamically cold Jupiter is $\sim$75 Myr. This is related to the longer accretion timescales due to distant embryos as discussed above.

The timing of the last giant impact is in general earlier than the $t_{90}$ values listed in Table \ref{growthtimes}. This suggests that accretion following the last impact was substantial.  This is evident in the high values for $M_{veneer}$, which are typically around 20\%, much larger than the inferred value for the Earth, and larger than in the simulations of \citet{OBrien2006bIcar}, which find median values of $\sim$1\% and $\sim$10\% depending on the excitation of Jupiter's orbit.

\begin{figure}[]
\centering
\begin{minipage}[b]{0.5\linewidth}
\centering
\includegraphics[width=3.25in]{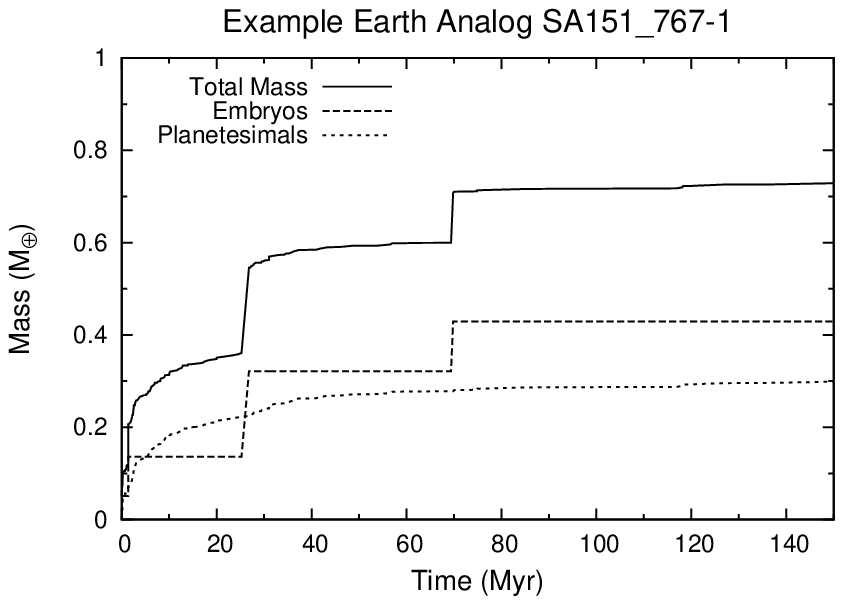}
\end{minipage}%
\begin{minipage}[b]{0.5\linewidth}
\centering
\includegraphics[width=3.25in]{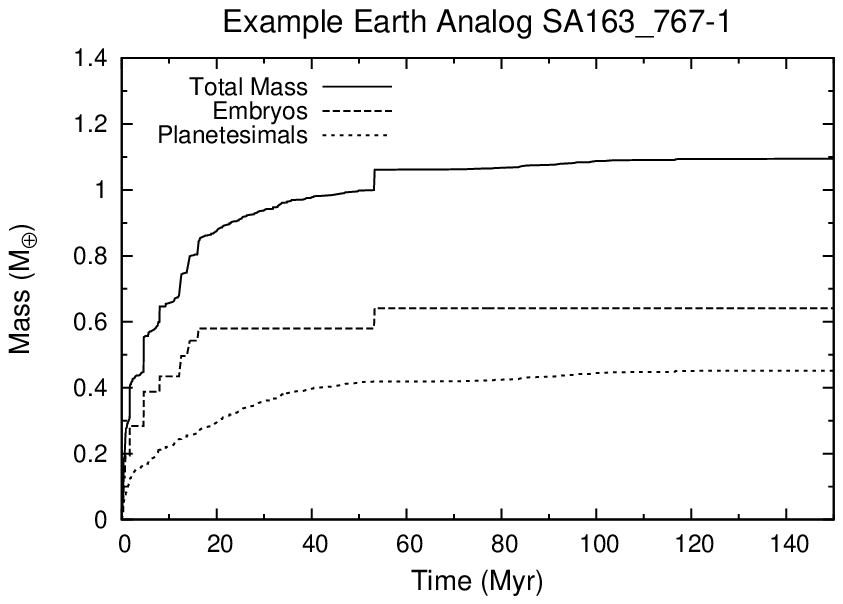}
\end{minipage}
\figcaption{Growth curves for two different Earth analogues from Table \ref{lastimpactstbl}, showing the total planet mass, as well as embryo and planetesimal contributions as a function of time. \label{eafig2}}
\end{figure}

Table \ref{eatable} lists all planets larger than 0.5 $\mathrm{M_{\earth}}$ in our simulations that are impacted by a body of at least an embryo mass later than 20 Myr (which we very broadly term an ``Earth Analogue'' here).  Growth curves for two of these planets are shown in Figure \ref{eafig2}.  Only 8 out of the 27 planets larger than 0.5 $M_{\earth}$ experience such an impact.  Excluding the three cases with late veneer values around 10\%, that leaves 5 out of 27 planets larger than 0.5 $\mathrm{M_{\earth}}$ that approximately satisfy most of the restrictive constraints for both the timing and nature of the last giant impact on the Earth.

\subsection{Planetesimal Accretion and Water Delivery}

Table \ref{impstatstbl} shows statistics related to the accretion of planetesimals in our simulations, broken down by simulation group and planet mass.  $f_{ptsml}$ gives the median planet mass fraction that consists of planetesimals, and for all simulations and final planet masses, roughly half of the accreted mass arrives as individual planetesimals (rather than embryos or planetesimals that have already been accreted by another embryo). $t_{50,ptsml}$ gives the timescale for a planet to accrete half of the total planetesimals that it will eventually accrete.  

$f_{belts}$, $t_{50,belts}$, $f_{disk}$ and $t_{50,disk}$ give the corresponding values for the ``belts'' and ``disk'' populations of primitive planetesimals.  Recall from Sec.~\ref{simulations} that we defined the masses of planetesimals in these two populations assuming that either one or the other supplied all of the primitive bodies in the asteroid belt.  Hence, the total water delivered to a planet will not be the sum of $f_{belts}$ and $f_{disk}$, but rather somewhere between the two.

We can calculate the overall accretion efficiencies for the primitive planetesimals, which we define as the total mass of primitive planetesimals in the final terrestrial planets divided by the total mass of primitive material available at the start of the simulation.  Accretion efficiencies are 59.1\% and 10.5\% for the belts and disk planetesimals in the *563 simulations (which includes all primitive planetesimals that have q$<$1.5 AU) and 51.7\% and 7.6\% in the *767 simulations (which includes all primitive planetesimals having q$<$2.0 AU).  The slightly lower values for the simulations with the $q<$2.0 AU cutoff suggest that relatively fewer planetesimals in the $q$ = 1.5--2.0 AU range are accreted by the terrestrial planets, compared to those with $q<$1.5 AU.

\begin{figure}[]
\centering
\begin{minipage}[b]{0.5\linewidth}
\centering
\includegraphics[width=3.25in]{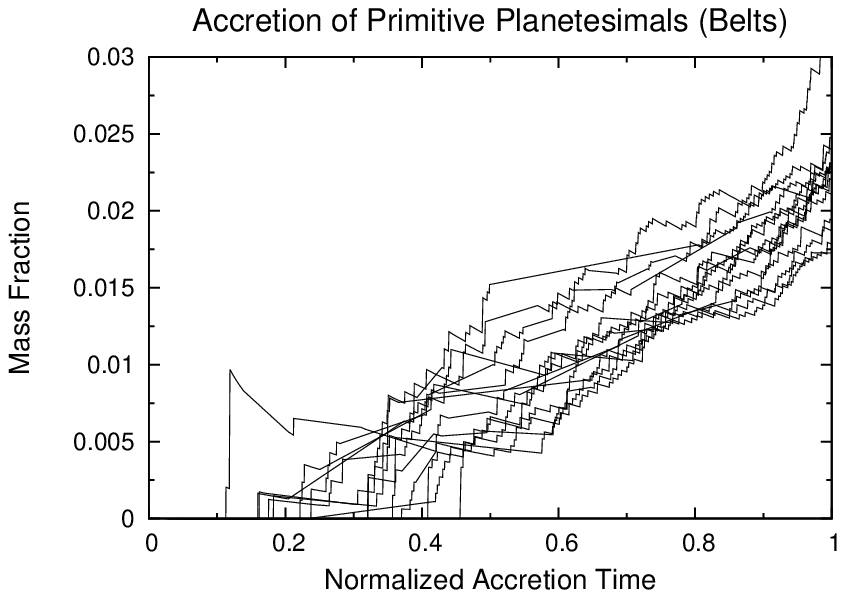}
\end{minipage}%
\begin{minipage}[b]{0.5\linewidth}
\centering
\includegraphics[width=3.25in]{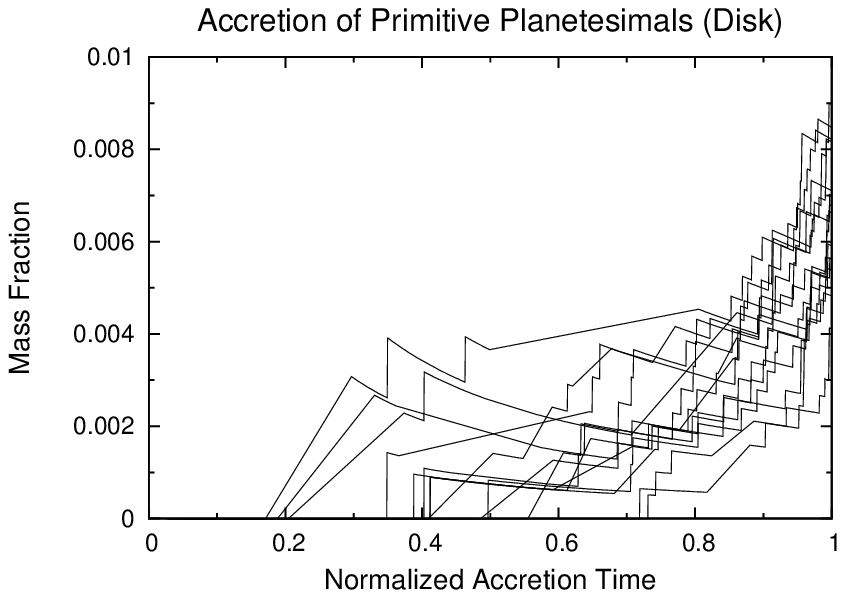}
\end{minipage}
\figcaption{Accretion curves for all planets larger than 0.75 $\mathrm{M_{\earth}}$, showing the mass fraction of primitive planetesimals from the `belts' and `disk' populations as a function of the normalized time, which is the mass of the planet (immediately following impact) divided by the final mass of the planet at the end of the simulation.  The fact that the curves all increase with time indicates that the primitive material, and hence water, arrives relatively late in the accretion process (see also Table \ref{impstatstbl}).  \label{accretcurves}}
\end{figure}

$t_{50,belts}$ and $t_{50,disk}$ are much larger than both $t_{50}$ and $t_{50,ptsml}$, meaning that the primitive planetesimals arrive significantly later in the accretion process than both the embryos and the non-primitive planetesimals from the inner Solar System.  Figure \ref{accretcurves} shows this in a different way, plotting $f_{belts}$ and $f_{disk}$ for all planets whose final mass is larger than 0.75 $\mathrm{M_{\earth}}$ vs.~the ``normalized accretion time,'' which we define here as the mass of the planet immediately following the impact, as a fraction of its final mass. Equal accretion per unit time would be a horizontal line on these plots, so the fact that the curves in Figure \ref{accretcurves} all increase with increasing time means that the primitive planetesimals preferentially arrive late in the planets' accretion.  This does not mean, however, that all of the water-bearing material is accreted during a short phase at the very end of accretion, for instance, after the Moon-forming event. Instead, water delivery is still a relatively gradual process, with the planets continuing to experience significant growth while accreting water-rich planetesimals.  The arrival of this primitive material later in accretion process, when the planet is larger, will make it easier for the water and volatiles that may be delivered by those impactors to be retained by the planet.

\begin{figure}[]
\centering
\includegraphics[width=5.5in]{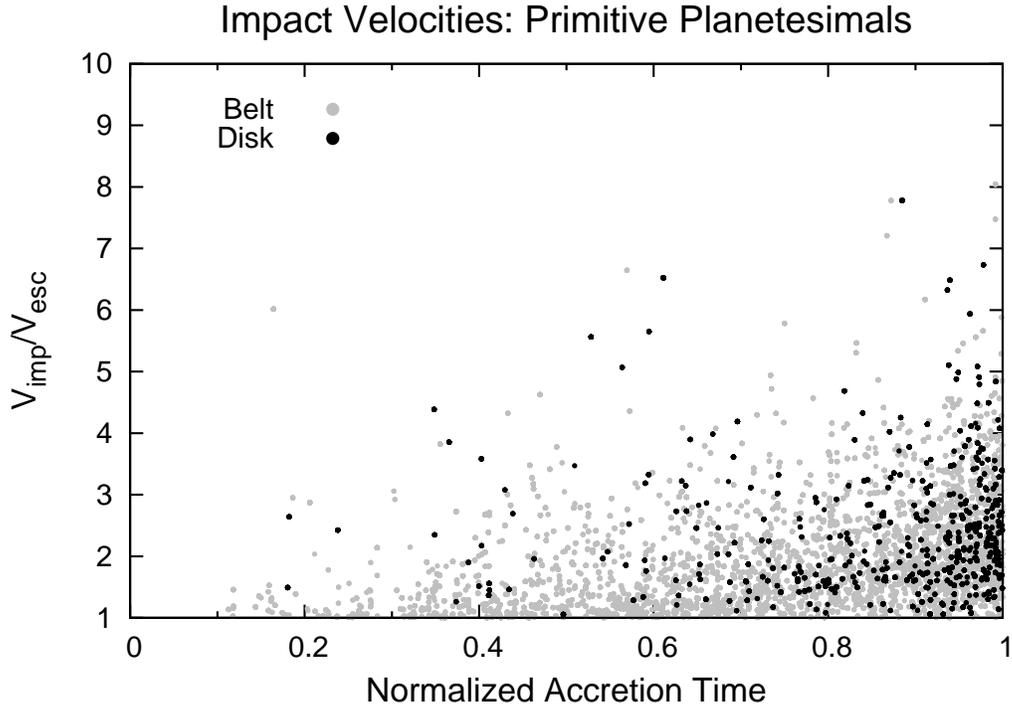}
\figcaption{Impact velocities onto all planets larger than 0.75 $\mathrm{M_{\earth}}$ in the simulations by primitive planetesimals from the `belts' and `disk' populations.  The impact velocity is normalized to the escape velocity of the planet at the time of impact and plotted vs.~normalized time, which is the mass of the planet (immediately following impact) divided by the final mass of the planet at the end of the simulation.  Median values of $V_{imp}/V_{esc}$ are 1.72 for the belt particles and 2.10 for the disk particles.  \label{vimpfig}}
\end{figure}

The impact velocity of these primitive planetesimals is also important for determining how much of the water that is delivered may be retained following the impact.  Figure \ref{vimpfig} shows $V_{imp}/V_{esc}$, the impact velocity as a fraction of the instantaneous escape velocity, plotted vs.~the normalized accretion time.  Median values of $V_{imp}/V_{esc}$ are 1.72 for the belt particles and 2.10 for the disk particles.
According to the simulations of  \citet{deNiem2012Icar}, vaporization of volatiles during impact can account for roughly 20\% volatile loss for a planetesimal impacting at $\sim$2$\times V_{esc}$. Thus, the accretion of the primitive planetesimals in our simulations would not be perfect, but the majority of the impacts are in a relatively efficient regime.

The total mass of water in the Earth's crust, oceans, and atmosphere is estimated to range from $2.8\times10^{-4}$ $\mathrm{M_{\earth}}$ \citep{Lecuyer1998ChemGeo}.  The amount of water in the mantle is more uncertain, and has been estimated to be $0.8-8\times10^{-4}$ $\mathrm{M_{\earth}}$ \citep{Lecuyer1998ChemGeo} up to $2 \times10^{-3}$ $\mathrm{M_{\earth}}$ \citep{Marty2012EPSL}.  Even more water, perhaps 10--50 Earth oceans, could have potentially been contained in the primitive mantle \citep{Dreibus1989OEPSA, Abe2000OEM, Righter1999EPSL}, although that amount has not been positively determined.  Hence, a reasonable lower limit for the amount of water that must be delivered to the Earth is $5\times10^{-4}$ $\mathrm{M_{\earth}}$.  Larger values would still be consistent, given the uncertainty in the water content of the primitive mantle.

We assume that the primitive bodies in our simulations, the belts and disk particles, have a water mass fraction of 10\% (0.1) by mass, consistent with water-rich carbonaceous chondrite meteorites.  This value is probably conservative given that asteroids known as main-belt comets appear to have a much larger water content than this \citep{Jewitt2012AJ}. In addition, water vapor has been discovered coming from Ceres \citep{Kuppers2014Nat} and water ice has been observed on the surface of Themis \citep{Campins2010Nat, Rivkin2010Nat}, suggesting that the water content of primitive asteroids, small and large, is substantially larger than that recorded in meteorites. We remind the reader that in meteorites only the water bound to the silicates can be found, all the water ice having been lost, whereas on asteroids water ice itself has been detected.  

The water content of the final planets would then be at least 0.1 times the fraction of the planet consisting of primitive material ($f_{belts}$ or $f_{disk}$ from Table \ref{impstatstbl}).  We find that planets larger than 0.75 $\mathrm{M_{\earth}}$ would have average water mass fractions of $2.3 \times10^{-3}$ if the water were delivered by the belts population, or $7 \times10^{-4}$ if it were delivered from the disk population, both of which exceed the lower limit of $5\times10^{-4}$ established above (4.6 and 1.4 times larger, respectively). The total amount of water accreted by the planets would lie between these values as the real population of primitive bodies delivering the water would be a combination, not a sum, of the belts and disk populations.  Note that even if the water content of primitive planetesimals was several tens of percent rather than the conservative 10\% value assumed here, the amount of delivered water would still be consistent with the range of current estimates, given possible volatile loss during impacts and the fact that 10--50 Earth oceans of water could have potentially been contained in the Earth's primitive mantle, as discussed above.

Previous simulation including a full disk of material would typically assign water mass fraction values as a function of initial heliocentric distance, increasing with distance. In these works water was delivered by embryos and planetesimals initially exterior to 2.5~AU, but interior to Jupiter \citep{Morbidelli2000MAPS,Raymond2004Icar,OBrien2006bIcar,Raymond2007AsBio,Raymond2009Icar,Izidoro2013ApJ}. With Jupiter on a circular orbit, these simulations delivered roughly $1.5\times10^{-2}$ $\mathrm{M_{\earth}}$ of water \citep{OBrien2006bIcar}. The reason for the substantially higher water delivery in many of these works was partly due to the fact that the bodies starting on initial orbits interior to the orbit of Jupiter have higher collision probabilities with the terrestrial planets than bodies originating beyond Jupiter.  Also, much of the water was delivered by large Mars-mass embryos originating from the outer asteroid belt region, and there are no such large primitive embryos included in the simulations presented here.  If the belts or the disk in our simulations originally contained a significant number of water-rich planetary embryos, there is a chance that one of them might have hit a planet and delivered its entire water budget.  In this sense, our estimate should be regarded as a lower limit.

On the other hand, the geochemistry of the Earth argues that carbonaceous chondritic material contributed at most only 2\% of the Earth's mass \citep{Marty2012EPSL}. If losses during impacts are no more that 50\%, this argues against the impact of a large (on the order of 0.1 $\mathrm{M_{\earth}}$) primitive embryo with the Earth. Small embryos (with masses not exceeding 0.025 $\mathrm{M_{\earth}}$) or bodies intermediate in mass between planetesimals and embryos (a `super-Ceres'), though, would satisfy the constraints.  In principle one could envision a scenario of stochastic delivery of water to the planets, similar to that proposed by \citet{Bottke2010Sci} to explain the difference in late veneer between the Earth and the Moon, but here contributing a larger mass and not necessarily restricted to the post Moon-formation era. In absence of strong constraints, however, this scenario cannot be firmly supported.

\section{Summary and Implications}

The work here explores in further detail the Grand Tack scenario as described in \citet{Walsh2011Nat}.  The inward-then-outward migration of Jupiter in that scenario provides a unique mechanism for truncating the disk of embryos and planetesimals in the inner Solar System, which has been shown by \citet{Hansen2009ApJ} to lead to a more realistic distribution of terrestrial planets, and also for delivering primitive planetesimals to the terrestrial planets from beyond the orbit of Jupiter.  Using the same initial conditions as \citeauthor{Walsh2011Nat}, we quantify the amount of water that may be delivered to the terrestrial planets by that mechanism, and analyze the statistical distributions of the properties of terrestrial planet systems formed in this model.

The major advantages of terrestrial planet formation starting from a truncated disk, or annulus \citep{Hansen2009ApJ}, were confirmed in this work. The distribution of planet mass as a function of semi-major axis in our simulations is broadly consistent with the terrestrial planets of our Solar System, matching the results found previously \citep{Hansen2009ApJ,Walsh2011Nat}.  In particular, the truncation of the disk at $\sim$1 AU leads to a high degree of radial mass concentration.  The dynamical excitation of these systems is also generally comparable to the current terrestrial planets.

The total amount of primitive, water-bearing material accreted by the planets is found to be on order 1--2\% of their total mass, and it tends to arrive during the second half of the accretion process, when it is more likely to be retained during collisions.  Using a conservative assumption that the primitive planetesimals have a water mass fraction of 10\%, consistent with the water content of some carbonaceous chondrite meteorites but significantly lower than the water content inferred for some primitive asteroids known as ``main-belt comets,'' and assuming that most of the water is retained by the planets during collisions, we find that the final planets have water mass fractions comparable to that estimated for the Earth today. 

Mars-like planets are often formed in our simulations, as individual embryos or low-mass planets that are scattered and isolated from the rest of the embryos and planetesimals.  However, those that do form at roughly the right mass generally take too long to form, compared to radiometric dating that suggests Mars formed within 2--10 Myr \citep[eg.][]{Nimmo2007Icar, Dauphas2011Nat}.  A better fit may be achieved by starting with embryos that are close to a Mars mass to begin with, such that a single embryo can be scattered onto a Mars-like orbit and reach a Mars mass without accreting further embryos.  Similarly, we are not able to produce reasonable Mercury analogues.  This may potentially be remedied by varying the embryo mass and/or surface density profile at the inner edge of the embryo/planetesimal disk, although the effects of giant impacts may have also played a role in stripping away much of Mercury's mantle.  These issues are currently being explored in further detail with simulations using a much larger range of initial embryo and planetesimal distributions \citep[see][]{Jacobson2014Sub}.  

We find that with the initial conditions used here, accretion timescales are relatively rapid compared to radiometric chronometers.  Large late impacts on Earth-mass planets, as would be needed to form the Moon, are not commonplace.  Most happen before 20 Myr, and are followed by the accretion of a fairly large amount of material, much more massive than that usually inferred from the geochemical ``late veneer'' of highly siderophile elements in the Earth's mantle.  Of the 27 planets larger than 0.5 $\mathrm{M_{\earth}}$ formed in all simulations, only 5 have late enough giant impacts and small enough late veneers to match constraints.  \citet{Jacobson2014Nat} have shown that the initial embryo mass and embryo/planetesimal mass ratio have a strong effect on the timing of the last giant impact and the mass of the late veneer, suggesting that a wider range of initial conditions must be explored in order to achieve a better match to all properties of the terrestrial planets.

\section*{Acknowledgments}

D.~P.~O'Brien was supported by grant NNX09AE36G from NASA's Planetary Geology and Geophysics research program. K.~J.~Walsh was supported by the NASA Lunar Science Institute (Center for Lunar Origin and Evolution at the Southwest Research Institute in Boulder, Colorado), NASA Grant NNA09DB32A.  D.~P.~O'Brien and A.~Morbidelli acknowledge European Research Council (ERC) Advanced Grant ``ACCRETE'' for support (contract number 290568).  We thank the two reviewers for their helpful comments and suggestions.

\clearpage


\clearpage

\clearpage

\clearpage

\clearpage

\clearpage


\begin{table}[]
\begin{center}
\textbf{Excitation and Radial Concentration of Final Systems}
\end{center}
\vspace{-0.2in}
\begin{center}
\begin{tabular}{c|cccc}
\hline 
\hline 
& $S_c$ & & $S_d$ & \\
& median & (range) & median & (range) \\
\hline 
SA15 & 88.6 & (68.7 to 105.3) & -0.00061 & (-0.00040 to -0.0032) \\
SA16 & 71.2 & (53.0 to 95.0) & -0.0017 & (-0.00098 to -0.0041) \\
Sol Sys & 89.9 & & -0.0018 & \\
\hline
\end{tabular}
\end{center}
\caption{Radial mass-concentration factors $S_c$ and normalized angular momentum deficits $S_d$ of the final terrestrial planet systems, using orbital elements averaged over the last 1 Myr of each simulation.  $S_c$ and $S_d$ for the Solar System's terrestrial planets are given for comparison, from \citet{Chambers2001Icar}.  High values of $S_c$ indicate a large degree of mass concentration and small values of $S_d$ (which is always negative) indicate low dynamical excitation. \label{amdtable}}
\end{table}

\clearpage

\begin{table}
\begin{center}
\textbf{Growth Timescales of Planets}
\end{center}
\vspace{-0.2in}
\begin{center}
\begin{tabular}{c|ccccc}
\hline
\hline
& & & $t_{50}$ (Myr) & $t_{90}$ (Myr) \\
& $M_{planet}$ & $N$ & median (range) & median (range) \\
\hline
 & $2 \ \mathrm{Emb} - 0.5 \ \mathrm{M_{\earth}}$ & 8 & 0.7 (0.2--11.2) & 27.3 (9.4--35.2) \\ 
SA15 & $0.5 - 0.75 \ \mathrm{M_{\earth}}$ & 6 & 2.9 (0.6--26.8) & 25.1 (22.4--69.8) \\  
 & $>0.75 \ \mathrm{M_{\earth}}$ & 7 & 2.0 (0.9--3.7) & 21.7 (15.8--28.1) \\ 
\hline
 & $2 \ \mathrm{Emb} - 0.5 \ \mathrm{M_{\earth}}$ & 13 & 1.6 (0.1--63.8) & 34.1 (18.4--63.8) \\
SA16 & $0.5 - 0.75 \ \mathrm{M_{\earth}}$ & 7 & 3.1 (1.5--11.9) & 29.0 (16.6--71.9) \\  
 & $>0.75 \ \mathrm{M_{\earth}}$ & 7 & 3.4 (1.3--13.4) & 27.4 (18.7--45.5) \\ 
\hline
 & $2 \ \mathrm{Emb} - 0.5 \ \mathrm{M_{\earth}}$ & 21 & 1.3 (0.1--63.8) & 33.3 (9.4--63.8) \\
All & $0.5 - 0.75 \ \mathrm{M_{\earth}}$ & 13 & 3.0 (0.6--26.8) & 25.4 (16.6--71.9) \\
 & $>0.75 \ \mathrm{M_{\earth}}$ & 14 & 2.1 (0.9--13.4) & 22.4 (15.8--45.5) \\
\hline
\end{tabular}
\end{center}
\caption{Growth timescales for planets in all simulations.  $M_{planet}$ is the mass of the planet, $N$ is the number of planets in the simulations that fall into that mass range, and $t_{50}$ and $t_{90}$ are the timescales necessary for a planet to reach 50\% and 90\% of its final mass.  There are 32 single-embryo planets not included in these statistics. \label{growthtimes}}
\end{table}

\clearpage

\begin{sidewaystable}
\begin{center}
\textbf{Statistics of Final Large Impacts}
\end{center}
\vspace{-0.2in}
\begin{center}
\begin{tabular}{c|cccccc}
\hline
\hline
& & & $t_{imp}$ (Myr) & $M_{imp}$ ($\mathrm{M_{\earth}}$) & $V_{\infty}$ (km/s) & $M_{veneer}$ \\
& $M_{planet}$ & $N$ & median (range) & median (range) & median (range) & median (range)\\
\hline
 & $2 \ \mathrm{Emb} - 0.5 \ \mathrm{M_{\earth}}$ & 8 & 4.9 (0.2--35.2) & 0.06 (0.05--0.12) & 2.54 (1.25--12.02) & 23.8\% (6.7--53.5) \\
SA15 & $0.5 - 0.75 \ \mathrm{M_{\earth}}$ & 6 & 5.9 (0.6--69.8) & 0.14 (0.09--0.19) & 1.15 (0.35--2.55) & 25.4\% (2.8--43.0) \\
 & $>0.75 \ \mathrm{M_{\earth}}$ & 7 & 5.6 (0.3--18.0) & 0.09 (0.06--0.12) & 2.38 (0.00--4.53) & 23.6\% (11.4--58.5) \\
\hline
 & $2 \ \mathrm{Emb} - 0.5 \ \mathrm{M_{\earth}}$ & 13 & 7.7 (0.0--68.8) & 0.04 (0.03--0.15) & 1.80 (0.34--8.21) & 20.8\% (0.8--66.9) \\
SA16 & $0.5 - 0.75 \ \mathrm{M_{\earth}}$ & 7 & 17.6 (9.2--71.9) & 0.05 (0.03--0.12) & 4.25 (1.46--8.64) & 8.4\% (0.9--26.0) \\
 & $>0.75 \ \mathrm{M_{\earth}}$ & 7 & 15.5 (3.5--147.3) & 0.05 (0.03--0.26) & 4.08 (0.00--18.41) & 12.0\% (0.0--31.5) \\
\hline
 & $2 \ \mathrm{Emb} - 0.5 \ \mathrm{M_{\earth}}$ & 21 & 6.3 (0.0--68.8) & 0.06 (0.03--0.15) & 2.12 (0.34--12.02) & 21.0\% (0.8--66.9) \\ 
All & $0.5 - 0.75 \ \mathrm{M_{\earth}}$ & 13 & 16.6 (0.6--71.9) & 0.11 (0.03--0.19) & 1.94 (0.35--8.64) & 12.6\% (0.9--43.0) \\
 & $>0.75 \ \mathrm{M_{\earth}}$ & 14 & 8.5 (0.3--147.3) & 0.07 (0.03--0.26) & 2.47 (0.00--18.41) & 18.2\% (0.0--58.5) \\
\hline
\end{tabular}
\end{center}
\caption{Statistics for final large impacts in our simulations, where a large impact is one in which the impactor is at least as large as a single embryo.  $M_{planet}$ is the mass of the planet, $N$ is the number of planets in the simulations that fall into that mass range, $t_{imp}$ is the time of the final large impact, $M_{imp}$ is the mass of the impactor, $V_{\infty}$ is the velocity of the impactor at infinity (ie.~not taking gravitational focusing into account) and $M_{veneer}$ is the total mass of planetesimals accreted after the final large embryo impact, given as a fraction of the total mass of the planet. There are 32 single-embryo planets not included in these statistics. \label{lastimpactstbl}}
\end{sidewaystable}

\clearpage

\begin{sidewaystable}
\begin{center}
\textbf{Planets with Late Final Impacts}
\end{center}
\vspace{-0.2in}
\begin{center}
\begin{tabular}{ccccccc}
\hline
\hline
Sim-Planet & $M_{planet}$ ($\mathrm{M_{\earth}}$) & a (AU) & $t_{imp}$ (Myr) & $M_{imp}$ ($\mathrm{M_{\earth}}$) & $V_{\infty}$ (km/s) & $M_{veneer}$ \\
\hline
SA151\_563-1 & 0.62 & 0.61 &  25.4 & 0.165 &  1.16 &  8.7\% \\
SA151\_767-1 & 0.73 & 0.62 &  69.8 & 0.108 &  1.42 &  2.8\% \\
SA162\_767-2 & 0.91 & 0.85 &  22.1 & 0.044 &  4.08 &  9.0\% \\
SA163\_563-2 & 1.04 & 0.76 & 147.3 & 0.028 & 18.41 &  0.0\% \\
SA163\_563-3 & 0.65 & 1.18 &  71.9 & 0.119 &  1.46 &  0.9\% \\
SA163\_767-1 & 1.09 & 0.64 &  53.3 & 0.062 &  6.74 &  3.0\% \\
SA164\_563-1 & 0.66 & 0.59 &  21.4 & 0.042 &  7.64 & 12.6\% \\
SA164\_767-1 & 0.59 & 0.59 &  66.7 & 0.031 &  8.64 &  3.0\% \\
\hline
\end{tabular}
\end{center}
\caption{Planets larger than 0.5 $\mathrm{M_{\earth}}$ with final large impacts occurring later than 20 Myr (these are broadly termed ``Earth Analogues'' in the text).  Planets are identified by simulation number and a planet number, where 1 is the closest planet to the Sun.  $M_{planet}$ is the mass of the planet, $a$ is the semimajor axis, $t_{imp}$ is the time of the final large impact, $M_{imp}$ is the mass of the impactor, $V_{\infty}$ is the velocity of the impactor at infinity (ie.~not taking gravitational focusing into account) and $M_{veneer}$ is the total mass of planetesimals accreted after the final large embryo impact, given as a fraction of the total mass of the planet. \label{eatable}}
\end{sidewaystable}

\clearpage

\renewcommand{\arraystretch}{0.67}

\begin{sidewaystable}
\begin{center}
\textbf{Statistics of Planetesimal Impacts}
\end{center}
\vspace{-0.2in}
\begin{center}
\begin{tabular}{c|ccccccccc}
\hline
\hline
& $M_{planet}$ & $N$ & $t_{50}$ (Myr) & $f_{ptsml}$ & $t_{50,ptsml}$ (Myr) & $f_{belts}$ & $t_{50,belts}$ (Myr) & $f_{disk}$ & $t_{50,disk} (Myr) $ \\
\hline
 & $2 \ \mathrm{Emb} - 0.5 \ \mathrm{M_{\earth}}$ & 8 & 0.7 & 54.5\% & 2.6 & 2.0\% & 17.1 & 0.6\% & 36.5 \\
SA15* & $0.5 - 0.75 \ \mathrm{M_{\earth}}$ & 6 & 2.9 & 51.7\% & 4.7 & 2.8\% & 15.7 & 0.7\% & 34.8 \\
 & $>0.75 \ \mathrm{M_{\earth}}$ & 7 & 2.0 & 54.1\% & 3.4 & 2.1\% & 10.0 & 0.6\% & 28.9 \\
\hline
 & $2 \ \mathrm{Emb} - 0.5 \ \mathrm{M_{\earth}}$ & 13 & 1.6 & 51.2\% & 4.6 & 2.3\% & 20.2 & 0.8\% & 33.0 \\
SA16* & $0.5 - 0.75 \ \mathrm{M_{\earth}}$ & 7 & 3.1 & 38.4\% & 6.2 & 2.3\% & 17.5 & 0.6\% & 20.6 \\
 & $>0.75 \ \mathrm{M_{\earth}}$ & 7 & 3.4 & 43.9\% & 7.2 & 2.3\% & 15.5 & 0.7\% & 45.4 \\
\hline
 & $2 \ \mathrm{Emb} - 0.5 \ \mathrm{M_{\earth}}$ & 11 & 2.9 & 53.6\% & 6.5 & 2.3\% & 20.2 & 0.6\% & 37.0 \\
SA*563 & $0.5 - 0.75 \ \mathrm{M_{\earth}}$ & 7 & 3.1 & 48.1\% & 5.6 & 2.6\% & 15.6 & 0.7\% & 27.3 \\
 & $>0.75 \ \mathrm{M_{\earth}}$ & 7 & 2.8 & 45.0\% & 6.4 & 2.1\% & 11.6 & 0.7\% & 37.3 \\
\hline
 & $2 \ \mathrm{Emb} - 0.5 \ \mathrm{M_{\earth}}$ & 10 & 1.0 & 50.8\% & 2.2 & 2.1\% & 18.4 & 0.9\% & 29.6 \\
SA*767 & $0.5 - 0.75 \ \mathrm{M_{\earth}}$ & 6 & 2.7 & 41.8\% & 4.7 & 2.5\% & 17.3 & 0.7\% & 25.1 \\
 & $>0.75 \ \mathrm{M_{\earth}}$ & 7 & 1.8 & 48.5\% & 6.2 & 2.3\% & 10.5 & 0.7\% & 35.5 \\
\hline
 & $2 \ \mathrm{Emb} - 0.5 \ \mathrm{M_{\earth}}$ & 21 & 1.3 & 51.8\% & 3.6 & 2.2\% & 19.7 & 0.7\% & 35.2 \\
All & $0.5 - 0.75 \ \mathrm{M_{\earth}}$ & 13 & 3.0 & 42.6\% & 5.0 & 2.6\% & 17.1 & 0.7\% & 27.3 \\
 & $>0.75 \ \mathrm{M_{\earth}}$ & 14 & 2.1 & 46.8\% & 6.3 & 2.3\% & 11.4 & 0.7\% & 36.4 \\
\hline
\end{tabular}
\end{center}
\caption{Statistics of planetesimal impacts in the simulations.  $M_{planet}$ is the mass of the planet, $N$ is the number of planets in the simulations that fall into that mass range, and $t_{50}$ is the median accretion timescale for a planet to reach 50\% of their final mass (from Table \ref{growthtimes}). $f_{ptsml}$ is the median mass fraction of planetesimals in the final planet, where we only count planetesimals that are directly accreted;~planetesimals that were first accreted by an embryo that then goes on to collide with the planet are counted as accreted embryo mass.  $f_{belts}$ and $f_{disk}$ are the median mass fractions of `primitive' planetesimal from the `belts' and `disk' populations that are incorporated into the final planets;~all mass from those regions is included, whether it arrives as a directly-accreted planetesimal or if it was accreted by another embryo prior to merging with the planet.  Generally over 90\% of the `belts' and `disk' planetesimals are directly accreted by a given planet rather than being accreted by another embryo first.  $t_{50,belts}$ and $t_{50,disk}$ are the median timescales to accrete half of the primitive planetesimals that will eventually end up in the planet.  The fact that $t_{50,ptsml}$, and especially $t_{50,belts}$ and $t_{50,disk}$, are larger than $t_{50}$ shows that planetesimals, and especially primitive planetesimals, arrive late in a planet's accretion (see also Figure \ref{accretcurves}).  There are 32 single-embryo planets not included in these statistics.  \label{impstatstbl}}
\end{sidewaystable}

\renewcommand{\arraystretch}{1.0}

\end{document}